\documentclass[american,english]{revtex4-2}
\usepackage[T1]{fontenc}
\usepackage[utf8]{inputenc}
\usepackage{amsmath}
\usepackage{amssymb}
\usepackage{graphicx}
\usepackage{geometry}
\makeatletter
\usepackage{amsmath}

\makeatother

\usepackage{babel}
\begin{document}
\global\long\def\ket#1{\left|#1\right\rangle }%

\global\long\def\bra#1{\left\langle #1\right|}%

\global\long\def\braket#1#2{\left\langle #1\left|#2\right.\right\rangle }%

\global\long\def\ketbra#1#2{\left|#1\right\rangle \left\langle #2\right|}%

\global\long\def\braOket#1#2#3{\left\langle #1\left|#2\right|#3\right\rangle }%

\global\long\def\mc#1{\mathcal{#1}}%

\global\long\def\nrm#1{\left\Vert #1\right\Vert }%

\title{Heisenberg Echo Symmetrization for Gate-Insertion Error Mitigation }
\begin{abstract}
Gate insertion (GI) quantum error mitigation replaces involutory gates
such as CZ and CX with $2j+1$ repetitions of the same gates to nominally
amplify noise by the same factor. Analytical and experimental results
show that this amplification scaling fails for non-commuting gate
noise, as typically encountered in realistic devices. Here we introduce
Heisenberg-echo symmetrization (HES), which combines standard GI with
an echo inspired by the Heisenberg representation to cancel the leading
GI error. In our simulations, HES reduces the absolute fidelity error
by one to two orders of magnitude compared with GI. Unlike the more
accurate Layered-KIK method, HES is compatible with any hardware and
requires only circuit-level access. It enables virtual noise scaling
across platforms, potentially reducing sampling overhead substantially.
Its agnostic noise amplification also provides full resilience to
temporal noise drift. For pure initial states, however, high-rank
observables can substantially increase the echo variance relative
to non-echoed circuits, whereas rank-one observables incur no additional
sampling overhead compared with GI or Layered-KIK. HES is therefore
particularly suited to rank-one or nearly rank-one observables.
\end{abstract}
\author{Raam Uzdin}
\email{raam@mail.huji.ac.il}

\affiliation{\selectlanguage{american}%
Fritz Haber Research Center for Molecular Dynamics, Institute of Chemistry,
The Hebrew University of Jerusalem, Jerusalem 9190401, Israel}

\maketitle
In recent years, quantum error mitigation (QEM) \citep{cai2022quantum}
has proved indispensable for extracting meaningful results from noisy
quantum computers \citep{berg2022probabilistic,kim2023evidence,kim2023scalable,haghshenas2024probing,liao2024MLqem,eddins2022doubling,ebadi2022quantum,Hemery2024quant,fracZNEvolume,haghshenas2026digital,Qedma2026Advantage}.
Unlike quantum error correction (QEC), which requires substantial
hardware overhead, QEM can often be implemented without additional
hardware. Its overhead instead arises from the increased runtime required
to compensate for the larger variance of mitigated expectation-value
estimates.

Although this overhead can remain acceptable for experiments involving
dozens of qubits or more, QEM is not a scalable solution for large-volume
circuits at present noise levels. Combining QEM with QEC, however,
can potentially provide both scalability and robustness. QEC can address
local, rapidly accumulating noise components, while QEM can target
error events that are difficult to handle with QEC, such as leakage
and correlated errors. Moreover, for a given fidelity target, QEM
can reduce the hardware overhead required by QEC. Proposals for integrating
QEC and QEM are discussed in Refs.  \citep{suzuki2022quantum,2021errorQEMforTgatesQEC,lostaglio2021error,Qedma2025ImportanceConcept,wahl2023ZNEforQEC,Qedma2025syndrome,QECQEMsoftInfo,Cai2025virtualChan,ZNEfitLogicalQubits2026,Cai2026EMPhysQQEC}.

Many QEM methods construct an inverse noise channel through post-processing
of results obtained from noisy variants of the target circuit. Some
methods, such as probabilistic error cancellation (PEC) \foreignlanguage{american}{\citep{temme2017ZNEPECorig,mari2021extending,PECtimeDrift}}
and machine-learning-based methods \citep{liao2024MLqem,kim2020neural,muqeet2024QLEAR},
require noise characterization to construct the inverse circuits,
whereas others do not. This distinction is crucial because methods
based on noise characterization are intrinsically sensitive to noise
drift, that is, changes in the noise parameters during an experiment.
In contrast, characterization-free methods can be fully resilient
to such drift, provided that the noise varies slowly compared with
the time required to execute a “protocol batch,” in which one shot
is performed for each noise-amplified circuit \citep{npjqiKIK}. This
advantage is especially relevant when mitigating strong noise, as
the increased sampling overhead prolongs the experiment, thereby increasing
the likelihood of substantial noise variations compared with the original,
unmitigated experiment.

Characterization-free methods currently rely on either noise amplification
or purification \citep{Koczor2021,huggins2021virtual,araki2025spacetimeVD}.
Purification requires additional hardware overhead and is limited
to specific types of noise. The dual-state purification method \citep{huo2022DualState}
does not involve a hardware overhead but is limited to ``first-order
mitigation''. We therefore focus on amplification-based methods.

The original proposal for zero-noise extrapolation (ZNE) suggested
implementing noise amplification through pulse stretching \citep{temme2017ZNEPECorig}.
This approach was subsequently demonstrated experimentally in superconducting
circuits in Refs. \citep{kandala2019error,kim2023scalable}. However,
pulse stretching is difficult to apply because it requires careful
calibration for each stretching factor. It also assumes that the effective
noise scales linearly with gate duration.

Gate insertion (GI), also known as identity insertion or digital ZNE,
was proposed to overcome these difficulties and enable straightforward
noise amplification without access to pulse-level control. For multiqubit
gates that are self-inverse in the absence of noise, the method replaces
each gate with an odd number of repetitions, $2j+1$, with the expectation
that the noise is amplified by the same factor. However, this scaling
has been shown analytically and experimentally to be incorrect when
the noise does not commute with the ideal gate \citep{npjqiKIK}.
Because of this non-commutativity, GI acts as a noise-distortion protocol
rather than a noise-amplification protocol. Consequently, the mitigated
result does not converge to the correct value even as the mitigation
order is increased.

Despite these findings, GI remains widely used. In some cases, it
serves as a reference for demonstrating the advantages of newly introduced
mitigation methods. Such comparisons are of limited value because
GI is known to be flawed and unreliable, with no convergence guarantees.
Its widespread use is driven by its simplicity of implementation and
broad platform compatibility: it requires only the ability to implement
CX or CZ gates. In practice, this makes GI compatible with essentially
any computing platform that supports the universal gate-set paradigm.
In this work, we aim to retain this simplicity while correcting the
fundamental flaw of GI. We also provide analytical expressions that
explain the limitations of our approach.

Reference \citep{npjqiKIK} showed that the flaw in GI can be corrected
by using a noisy “pulse-inverse” evolution in place of the self-inverse
operation used in GI. The resulting protocol, KIK, has analytical
convergence guarantees for moderate noise. A subsequent extension
improved the approach by introducing noise amplification in layers
\citep{LKIK}. The resulting Layered-KIK protocol is bias-free for
all practical purposes (see further details in \citep{LKIK}) even
under strong noise and is also compatible with mid-circuit measurements
and feedforward, that is, dynamic circuits. This compatibility is
important both for implementing a wider range of algorithms and for
integration with QEC protocols.

Nevertheless, KIK is not automatically and universally applicable
across hardware platforms. It can be implemented straightforwardly
with Mølmer--Sørensen gates in trapped ions and with parametric tunable-coupler
gates and cross-resonance gates in superconducting circuits. For other
implementations, such as adiabatic tunable-coupler gates, it remains
unclear how KIK can be implemented reliably. On other platforms, including
neutral atoms and optical systems, the feasibility of implementing
KIK has not yet been explored to the best of our knowledge. KIK therefore
does not offer the same universal compatibility or simplicity as GI.

Here, we retain the universality and gate-level implementation of
GI while eliminating its leading error term. For a general observable
and initial pure state the Heisenberg-echo symmetrization (HES) comes
with an increased sampling overhead. As a result this method is mostly
suited to observables which are close to being pure. 

\section*{The gate insertion method and its flaw}

Throughout this work, we represent quantum states in Liouville space
by vectorizing the density matrix into a column vector, denoted by
$\lvert\rho\rangle$ \citep{gyamfi2020fundamentals}. In this form,
the action of a quantum channel is represented by left multiplication
with the corresponding Liouville-space operator, providing a convenient
description of noisy quantum evolution.

For an ideal Hilbert-space evolution generated by a unitary operator
$U_{H}$, the associated Liouville-space operator is $U=U_{H}\otimes U^{*}_{H}$.
The noisy evolution operator can be written as
\begin{equation}
K=UN,
\end{equation}
where $N$ is the noise channel operator in Liouville space. As in
the GI protocol, we amplify the noise of multiqubit involutory gates,
i.e. gates whose ideal channels satisfy $U^{2}=I$ or, equivalently
$U^{2j+1}=U$. We further assume that the gate is Clifford. This restriction
does not impose a practical limitation as it includes the commonly
used two-qubit CX and CZ gates. Finally, we assume that the gate is
Pauli twirled, thereby converting coherent errors into stochastic
Pauli noise in the averaged channel. The gate insertion protocol aims
to leverage this property and amplify the noise using an odd number,
$2j+1$, of repetitions of $K$ 
\begin{align}
K^{2j+1} & =Ue^{\chi}[Ue^{\chi}Ue^{\chi}]^{j}\nonumber \\
 & =UN[e^{\bar{\chi}}e^{\chi}]^{j},
\end{align}
where $\chi=\ln N$ and $\bar{\chi}=U\chi U$. Here $\chi$ denotes
the exact logarithmic generator of the Pauli-twirled noise channel
and is not obtained by truncating a Magnus expansion as done in Ref.
\citep{LKIK}. Since $U$ is Clifford, $\bar{N}=UNU$ is also a Pauli
noise channel. Since all Pauli channels commute it holds that $[\chi,\bar{\chi}]=0$,
which leads to
\[
K^{2j+1}=Ue^{(j+1)\chi+j\bar{\chi}}=Ue^{(2j+1)\frac{\chi+\bar{\chi}}{2}+\frac{\chi-\bar{\chi}}{2}}.
\]
If we use the Taylor coefficients corresponding to mitigation order
m, a residual bias remains due to the terms that do not scale with
$2j+1$:

\begin{align}
\sum^{m}_{j=0}a^{(m)}_{j}(UN)^{2j+1} & =U+U\frac{\chi-\bar{\chi}}{2}+\frac{1}{2}U(\frac{\chi-\bar{\chi}}{2})^{2}+O(\nrm{\frac{\chi+\bar{\chi}}{2}}^{m+1})\nonumber \\
 & =U+\frac{1}{2}[U,\chi]+\frac{1}{2}U\frac{\chi-\bar{\chi}}{2}UU\frac{\chi-\bar{\chi}}{2}+O(\nrm{\frac{\chi+\bar{\chi}}{2}}^{m+1})\nonumber \\
 & =U+\frac{1}{2}[U,\chi]-\frac{1}{8}U([U,\chi])^{2}+O(\nrm{\frac{\chi+\bar{\chi}}{2}}^{m+1},\nrm{[U,\chi]}^{3}).\label{eq: GI Taylor}
\end{align}
Naturally, if $[U,\chi]=0$ , gate insertion implements perfect noise
amplification. This can occur, for example, when the noise is a global
depolarizing channel, where “global” here refers at least to the qubits
on which the gate acts. However, global depolarizing noise cannot
capture common physical noise mechanisms in quantum computers, such
as decoherence, decay (amplitude damping), and leakage. More generally,
the validity of the condition $[U,\chi]=0$ is platform dependent.
For example in CZ gates decoherence noise commutes with the gate but
amplitude damping does not \citep{malekakhlagh2025efficient}. In
contrast, in trapped ions decoherence noise does not commute with
the $R_{XX}$ Mølmer--Sørensen gate. As pointed out in Ref. \citep{npjqiKIK},
Eq. (\ref{eq: GI Taylor}) reveals the critical flaw of the GI protocol:
when $[U,\chi]\neq0$ , gate insertion deviates from the correct result
already at first order in $\chi$.

It should be noted that, for a specific circuit, observable, and initial
state, the contribution of $[U,\chi]$ to the expectation value can
accidentally vanish. However, it is generally impractical to know
in advance whether this is the case, as doing so would require explicitly
solving the noisy circuit. Moreover, for a particular example and
at a particular mitigation order, gate insertion may appear to perform
better than the KIK method if the contribution to the expectation
value arising from $\frac{1}{2}[U,\chi]$ has the opposite sign to
the contribution arising from $(\frac{\chi+\bar{\chi}}{2})^{m+1}$.
Such an accidental cancellation is not expected to persist as the
mitigation order m increases, since $\frac{1}{2}[U,\chi]$ is independent
of m, whereas $(\frac{\chi+\bar{\chi}}{2})^{m+1}$ vanishes as $m$
increases.

\section*{The Heisenberg echo and its symmetry}

In the Heisenberg representation, operators are propagated backward
in time, and the final expectation value is evaluated by taking the
inner product with the initial state. For unitary evolution in the
Schrödinger representation in Hilbert space, the expectation value
of an observable A is $\left\langle A\right\rangle =\mathrm{tr}[A\rho(t)]=\mathrm{tr}[AU_{H}\rho(0)U^{\dagger}_{H}]$
, where $U_{H}$ is the unitary evolution operator in Hilbert space.
Using the cyclic property of the trace, this can equivalently be written
as $\left\langle A\right\rangle =\mathrm{tr}[\rho(0)U^{\dagger}_{H}AU_{H}]=\mathrm{tr}[\rho(0)A(t)]$,
which corresponds to the Heisenberg representation.

This expression suggests an alternative interpretation in which $\rho(0)$
plays the role of the observable, while A plays the role of an “initial
state” that is propagated backward in time. Although A is Hermitian,
it is generally not a valid density matrix because it is not necessarily
positive and does not necessarily have unit trace. This can be remedied
by introducing a shifted and scaled version of $A$
\begin{equation}
\tilde{A}=g(A-\lambda_{min}I).
\end{equation}
Here, $\lambda_{min}$ is the minimum eigenvalue of $A$. The condition
$\mathrm{tr}(\tilde{A})=1$ yields $g=\frac{1}{\mathrm{tr}[A]-\lambda_{min}tr[I]}$
. With this choice, $\tilde{A}$ is a valid density matrix. The target
expectation value can therefore be written in terms of the initial
state $\tilde{A}$, the backward evolution $U^{\dagger}_{H}$, and
a deterministic scalar shift:
\begin{equation}
\left\langle A\right\rangle =\mathrm{tr}[\rho(0)U^{\dagger}_{H}(\frac{1}{g}\tilde{A}+\lambda_{min}I)U_{H}]=\frac{1}{g}\mathrm{tr}[\rho(0)U^{\dagger}_{H}\tilde{A}U_{H}]+\lambda_{min}.\label{eq: <A> g}
\end{equation}
We use the term “Heisenberg echo” for a circuit in which the time
evolution is reversed, i.e., echoed, while $\tilde{A}$ is used as
the initial state and $\rho(0)$ is used as the observable. 

It is insightful to rewrite the Heisenberg echo in Liouville space.
In the Schrödinger representation, for unitary evolution, we have
\begin{equation}
\left\langle \tilde{A}\right\rangle =\braOket{\tilde{A}}U{\rho_{0}}.
\end{equation}
Since $\left\langle \tilde{A}\right\rangle $ is real, taking the
Hermitian conjugate gives
\begin{equation}
\left\langle \tilde{A}\right\rangle =\left\langle \tilde{A}\right\rangle ^{\dagger}=\braOket{\rho_{0}}{\mc U^{\dagger}}{\tilde{A}},
\end{equation}
where the evolution is reversed and the observable and initial state
exchange roles, as in the Heisenberg representation. Note that here
we have used invariance under Hermitian conjugation rather than the
cyclic property of the trace. Thinking of the echo and conjugation
together as an operation, the expectation value has a “Heisenberg
symmetry,” since it is invariant under this operation. In what follows,
we use the Heisenberg echo for noisy circuits to correct the leading-order
error of the gate-insertion protocol.

For simplicity, consider a circuit consisting of two layers, $K_{a}=U_{a}N_{a}$
and $K_{b}=U_{b}N_{b}$. The expectation value is 
\[
\left\langle \tilde{A}\right\rangle =\braOket{\tilde{A}}{K_{b}K_{a}}{\rho_{0}}.
\]
Taking the Hermitian conjugate gives
\begin{align}
\left\langle \tilde{A}\right\rangle  & =\left\langle \tilde{A}\right\rangle ^{\dag}=\braOket{\rho_{0}}{K^{\dagger}_{a}K^{\dagger}_{b}}{\tilde{A}}\\
 & =\braOket{\rho_{0}}{N^{\dagger}_{a}U^{\dagger}_{a}N^{\dagger}_{b}U^{\dagger}_{b}}{\tilde{A}}.
\end{align}
If the noise channel is Hermitian, we obtain $N_{a(b)}U^{\dagger}_{a(b)}=K^{I}_{a(b)}$
where $K^{I}$ denotes the pulse-inverse operation defined in Ref.
\citep{npjqiKIK}. Therefore, for gates with Hermitian noise, $\left\langle A\right\rangle $
can be evaluated using a pulse-inverse Heisenberg echo (PI$\text{-}$HE)
\[
\left\langle \tilde{A}\right\rangle ^{\text{PI-HE}}=\braOket{\rho_{0}}{K^{I}_{a}K^{I}_{b}}{\tilde{A}}.
\]
The noise can be rendered Hermitian using pseudo-twirling \citep{2024PST}
for the non-Clifford elements and Pauli twirling for the Clifford
elements. At this point, $\left\langle \tilde{A}\right\rangle ^{\text{PI-HE}}$
is just an alternative protocol for evaluating $\left\langle \tilde{A}\right\rangle $.
In what follows, we use an echo with intentionally broken Heisenberg
symmetry to correct the main flaw of the GI protocol.

\section*{Mending the gate insertion protocol}

Consider a platform on which pulse-inverse implementations of the
multiqubit gates are either not possible or not accessible. As before,
we assume that these multiqubit gates are Clifford gates satisfying
$U^{2}_{\ensuremath{g}}=I$ , for example, CX and CZ. For simplicity,
we first consider a circuit containing only a single gate, $K_{g}=U_{g}N_{g}$.
Moreover, after applying Pauli twirling, $N_{g}=N^{\dagger}_{g}$
and the noise becomes Pauli noise. Noise amplification using gate
insertion then reads
\begin{align}
\left\langle \tilde{A}\right\rangle ^{\text{GI}}_{2j+1} & =\braOket{\tilde{A}}{K_{g}{}^{2j+1}}{\rho_{0}}\nonumber \\
 & =\braOket{\tilde{A}}{(U_{g}N_{g})^{2j+1}}{\rho_{0}}.
\end{align}
Next, since the pulse inverse of the multiqubit gate is not available,
we apply gate insertion also to the Heisenberg echo and obtain the
following protocol:
\begin{equation}
\left\langle A\right\rangle ^{\text{\text{HE-GI}}}_{2j+1}=\braOket{\rho_{0}}{(U_{g}N_{g})^{2j+1}}{\tilde{A}}.
\end{equation}
Since $K_{g}$ is used instead of $K^{I}_{g}$, the Heisenberg symmetry
is broken, and therefore $\langle\tilde{A}\rangle^{\text{HE-GI}}_{2j+1}$
is no longer guaranteed to equal $\langle\tilde{A}\rangle^{\mathrm{GI}}_{2j+1}$.
To understand the properties of $\left\langle \tilde{A}\right\rangle ^{\text{HE-GI}}$
we use $\left\langle \tilde{A}\right\rangle ^{\text{HE-GI}}=(\left\langle \tilde{A}\right\rangle ^{\text{HE-GI}})^{\dagger}$
which gives
\begin{align}
\left\langle \tilde{A}\right\rangle ^{\text{HE-GI}}_{2j+1} & =\braOket{\tilde{A}}{(N_{g}U_{g})^{2j+1}}{\rho_{0}}.
\end{align}
Since $(N_{g}U_{g})^{2j+1}=U_{g}(U_{g}N_{g})^{2j+1}U_{g}$ and $U_{g}(\chi-\bar{\chi})U_{g}=-(\chi-\bar{\chi})$
it follows that
\begin{align}
(N_{g}U_{g})^{2j+1} & =U_{g}e^{(2j+1)\frac{\chi+\bar{\chi}}{2}-\frac{\chi-\bar{\chi}}{2}}.
\end{align}
Hence, for the Heisenberg echo with gate insertion
\begin{equation}
\sum^{m}_{j=0}a^{(m)}_{j}(N_{g}U_{g})^{2j+1}=U_{g}-\frac{1}{2}[U_{g},\chi]-\frac{1}{8}U_{g}([U_{g},\chi])^{2}+O(\nrm{\frac{\chi+\bar{\chi}}{2}}^{m+1},\nrm{[U_{g},\chi]}^{3}).\label{eq: HE GI gate Taylor}
\end{equation}
Motivated by the fact that the $[U_{g},\chi]$ term in the Heisenberg
echo has opposite sign compared to the GI, we introduce our Heisenberg
echo symmetrization (HES) protocol. HES is the average of the ordinary
GI, and the Heisenberg echo with GI, i.e.
\begin{equation}
\left\langle \tilde{A}\right\rangle ^{\text{HES}}_{mit}=\frac{\left\langle \tilde{A}\right\rangle ^{\text{GI}}_{mit}+\left\langle \tilde{A}\right\rangle ^{\text{HE-GI}}_{mit}}{2}=\sum^{m}_{j=0}a^{(m)}_{j}\frac{\left\langle \tilde{A}\right\rangle ^{\text{GI}}_{2j+1}+\left\langle \tilde{A}\right\rangle ^{\text{HE-GI}}_{2j+1}}{2}.\label{eq: HEGI protocol}
\end{equation}
Each expectation value is mitigated separately using the Taylor coefficients.
For the single-gate example above, averaging Eqs. (\ref{eq: GI Taylor})
and (\ref{eq: HE GI gate Taylor}) gives 
\begin{align}
\frac{1}{2}\sum^{m}_{j=0}a^{(m)}_{j}(U_{g}N_{g})^{2j+1}+\frac{1}{2}\sum^{m}_{j=0}a^{(m)}_{j}(N_{g}U_{g})^{2j+1} & =U_{g}-\frac{1}{8}U_{g}([U_{g},\chi])^{2}\nonumber \\
 & +O(\nrm{\frac{\chi+\bar{\chi}}{2}}^{m+1},\nrm{[U_{g},\chi]}^{3}).\label{eq: 1L cgi}
\end{align}
The cancellation of the linear commutator terms $\pm\frac{1}{2}[U_{g},\chi]$
in HES is our main result.

While Layered-KIK has an $\Omega_{2}$ residual term when considering
a single layer, in the present approach the corresponding single-layer
residual term is quadratic in the noise and proportional to $\frac{1}{8}U_{g}([U_{g},\Omega_{1}])^{2}$.
To quantify the residual HES error more accurately, we must consider
multilayer circuits, as discussed next.

\subsection*{Accumulated errors from multiple gate insertion layers}

When there are $L$ layers, each containing a Clifford gate of the
type discussed above, the error bias, i.e., the expression that determines
the error in the infinite-order mitigation limit, contains the following
terms:
\begin{align}
K^{(\infty)}_{\text{HES}}-U_{id} & \cong-\frac{1}{8}\sum^{L}_{k=1}U_{L}\ldots U_{k}[U_{k},\chi_{k}]^{2}U_{k-1}\ldots U_{1}\nonumber \\
+ & \frac{1}{4}\sum^{L}_{k>l}U_{L}\ldots U_{k+1}[U_{k},\chi_{k}]U_{k-1}\ldots U_{l+1}[U_{l},\chi_{l}]U_{l-1}\ldots U_{1}\nonumber \\
+ & O(\nrm{[U_{k},\chi_{k}]}^{3}).\label{eq: error layers}
\end{align}
The first sum contains the layer-local quadratic terms that appear
in Eq. (\ref{eq: 1L cgi}), whereas the second sum contains cross-layer
terms. Neither type of term is cancelled by the Heisenberg echo, since
the echo changes the signs of both commutators simultaneously.

If some layers are treated using pulse inverses, these layers should
be omitted from both sums, and the local second-order Magnus terms
$\Omega^{(k)}_{2}$ from each KIK layer should be added, as described
in Ref. \citep{LKIK}. The first sum in Eq. (\ref{eq: error layers})
is analogous to the $\Omega^{(k)}_{2}$ error in Layered-KIK. The
cross-layer terms in the second sum, on the other hand, have no counterpart
in the Layered-KIK method. The reason is that the Layered-KIK cross
terms scale as $(2j+1)^{2}$ and are therefore removed at mitigation
order two or higher. 

Another important difference from Layered-KIK is that the accumulated
error arising from the $\Omega^{(k)}_{2}$ terms can be systematically
reduced by using thinner layers, and these layers can be thinner than
a two-qubit gate. In HES, this is not possible because the layers
must contain gates satisfying $U^{2j+1}=U$. Thus, we find that
\begin{equation}
\left\langle \tilde{A}\right\rangle ^{\text{HES}}_{mit}=\left\langle \tilde{A}\right\rangle _{ideal}+O(\nrm{\frac{\chi+\bar{\chi}}{2}}^{m+1},\frac{1}{8}\nrm{[U_{k},\chi_{k}]}^{2},\frac{1}{8}\nrm{[U_{k},\chi_{k}]}\nrm{[U_{l},\chi_{l}]}),
\end{equation}
where summation over $k$ and $l$ is implied. Importantly, both the
cross-layer terms and the quadratic terms in $[U_{k},\chi_{k}]$ are
already present in the GI protocol in the same form. HES does not
modify these contributions; it cancels the leading linear term.

These limitations do not necessarily imply that HES is restricted
to weak-noise regimes. Even when the noise is strong, if most of the
gate noise commutes with the unitary part of the gate, the expression
in Eq. (\ref{eq: error layers}) can remain small. 

\subsection*{Sampling overhead}\label{sec:Sampling-overhead}

In this section, we study the nontrivial sampling overhead that may
arise when using the HES method. We set aside the effect of using
the gate instead of its inverse, since a more significant effect arises
even when $K^{\dagger}$ can be implemented exactly. Writing again
the expectation value of $A$ in its forward and Heisenberg-echo forms,

\begin{align}
\left\langle A\right\rangle  & =\frac{1}{g}\bra{\tilde{A}}K\ket{\rho_{0}}+\lambda_{min},\\
\left\langle A\right\rangle _{HE} & =\frac{1}{g}\bra{\rho_{0}}K^{\dagger}\ket{\tilde{A}}+\lambda_{min},
\end{align}
and using the notation $\mu\circeq\bra{\tilde{A}}K\ket{\rho_{0}}$
we obtain
\begin{align}
\mathrm{Var}(A) & =\frac{1}{g^{2}}[\bra{\tilde{A}^{2}}K\ket{\rho_{0}}-\bra{\tilde{A}}K\ket{\rho_{0}}^{2}]\nonumber \\
 & =\frac{1}{g^{2}}[\bra{\tilde{A}^{2}}K\ket{\rho_{0}}-\mu^{2}],\\
\mathrm{Var}(A)_{HE} & =\frac{1}{g^{2}}[\bra{\rho^{2}_{0}}K^{\dagger}\ket{\tilde{A}}-\bra{\tilde{A}}K\ket{\rho_{0}}^{2}]\nonumber \\
 & =\frac{\mu(1-\mu)}{g^{2}}.\label{eq: varAhe}
\end{align}
The factor $1/g^{2}$ can be very large. For example, for a Pauli
observable, $1/g^{2}=2^{2n}$ where $n$ is the number of qubits.
However, the variance of a Pauli observable is bounded by one, so
$\bra{\tilde{A}}K\ket{\rho_{0}}$ must be small, and scale at most
as $O(g^{2})$ to satisfy the bound.  

To obtain Eq. (\ref{eq: varAhe}), we have used the fact that the
initial state is pure, $\rho^{2}_{0}=\rho_{0}$. Therefore, knowing
the expectation value of the forward evolution allows us to directly
evaluate the exact variance of the Heisenberg echo. For a Pauli observable
$\mu=2^{-n}(\left\langle A\right\rangle +1)$ and we get that $\mathrm{Var}(A)_{HE}\simeq2^{n}(\left\langle A\right\rangle +1)$
from which we conclude that the HES sampling overhead quickly diverge
with the number of qubits when $A$ is highly ``mixed''.

One can show that, for a pure initial state, $\mathrm{Var}(A)_{HE}\ge\mathrm{Var}(A)$
with equality when $A=\tilde{A}=\ketbra{\psi}{\psi}$, i.e., when
$A$ is also pure. To obtain a more explicit quantitative relation
between the two variances, we consider their difference and find
\begin{equation}
\mathrm{Var}(A)_{HE}-\mathrm{Var}(A)\le\frac{1}{g^{2}}[1-\mathrm{tr}(\tilde{A}^{2})],
\end{equation}
where we have used the inequality $\mathrm{tr}AB\le\nrm A_{op}\mathrm{tr}B$
for any two positive matrices $A$ and $B$. This bound shows that
the difference between $\mathrm{Var}(A)_{\mathrm{HE}}$ and $\mathrm{Var}(A)$
becomes negligible when $1-\mathrm{tr}(\tilde{A}^{2})\ll g^{2}\mathrm{Var}(A)$.
Thus, the additional variance introduced by the Heisenberg echo is
small when $\tilde{A}$ is sufficiently close to a pure state.

\section*{Numerical results}

To illustrate the role of HES, we construct random seven-qubit circuits
in which the two-qubit gates are CZ gates and the single-qubit gates
are random single-qubit unitaries. The locations of the CZ gates are
chosen randomly, and the circuit depth is seven CZ layers. The initial
state is $\ket{0000000}$, and the observable is $A=\ketbra{\psi^{id}_{f}}{\psi^{id}_{f}}$
where $\ket{\psi^{id}_{f}}$ is the ideal final state. Therefore,
$\left\langle A\right\rangle $ is equal to the fidelity, i.e., the
overlap with the ideal final pure state.

\begin{figure}
\begin{centering}
\includegraphics[width=16cm]{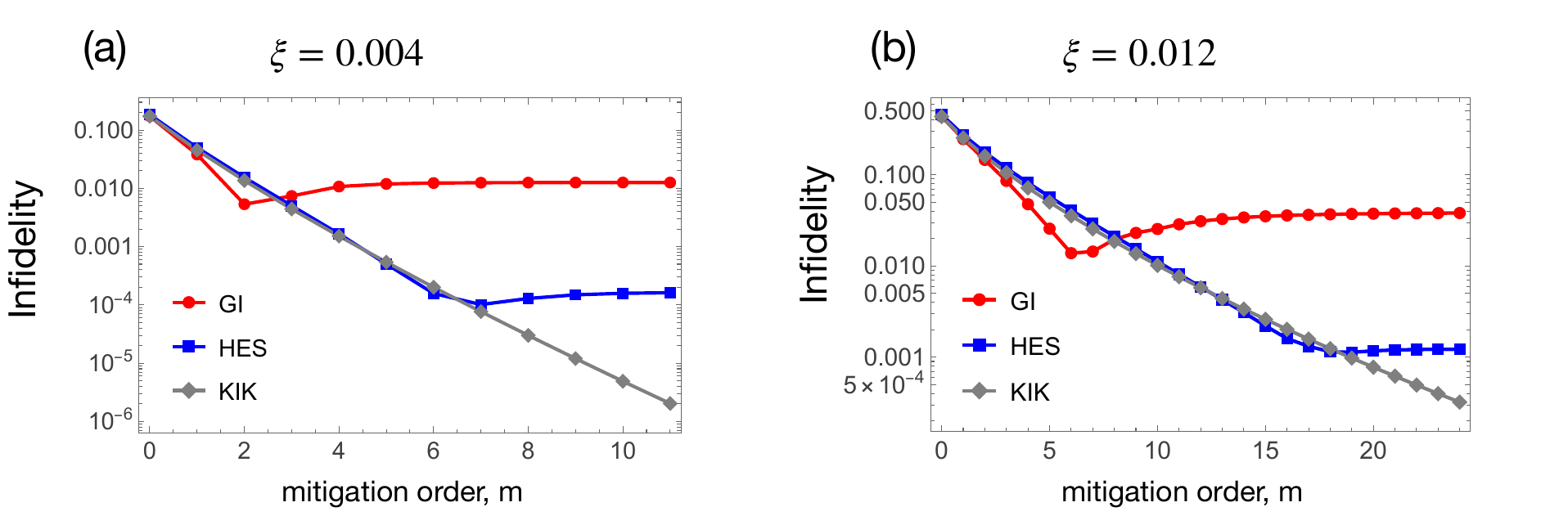}\caption{Mean infidelity as a function of the mitigation order $m$. The results
were averaged over 50 random seven-qubit circuits of depth seven (see
main text). The Heisenberg echo (HES-blue) outperforms the standard
gate insertion (GI-red) protocol. In the weak noise case (a) the HES
fidelity for large m is 80 times better than that of GI and in the
strong noise case (b) the improvement factor is 30. As expected from
theory, the pulse-inverse-based KIK method is more accurate than HES,
but HES has the same implementation simplicity and universal compatibility
as GI. }\label{fig: 1}
\par\end{centering}
\end{figure}

In our simulations, only the CZ gates are noisy. Each of the two qubits
participating in a CZ gate independently undergoes a $\sigma_{x}$
error with probability $\xi$ and a $\sigma_{y}$ error with probability
$\xi$. Figure \ref{fig: 1} shows the error for GI (red), HES (blue),
and KIK (grey) as a function of the mitigation order for $\xi=0.004$
(Figure \ref{fig: 1}a) and $\xi=0.012$ (Figure \ref{fig: 1}b).
The absolute error, $|\langle A\rangle_{\mathrm{ideal}}-\langle A\rangle_{\mathrm{mit}}|=1-\langle A\rangle_{\mathrm{mit}}$
was averaged over 50 random circuits. Mitigation order zero corresponds
to the error of the original circuit without mitigation.

In the weak-noise case, Fig. \ref{fig: 1}a, the final fidelity at
large mitigation order (m) is improved by a factor of 80 with HES
compared with GI. In the strong-noise case, Fig. \ref{fig: 1}b, the
improvement factor is 30. Virtual noise scaling can reduce the sampling
overhead by several orders of magnitude \citep{uzdin2026VNS}.

It is also observed that the HES curves closely follow the KIK curves
until the residual HES terms in Eq. (\ref{eq: error layers}) become
significant. The local decrease of the GI error below that of KIK
at $m=2$ in Fig. \ref{fig: 1}a and at $m=5$ in Fig. \ref{fig: 1}b
results from an accidental cancellation between the contribution to
$\langle A\rangle$ arising from the GI error term $[U,\chi]$ and
the contribution arising from the residual Taylor term.

When a projector onto a random pure state is chosen as the observable,
both the ideal and noisy expectation values are typically close to
zero. As a result, the absolute differences between HES, GI, and KIK
can also be small. Nevertheless, the advantage of HES over GI becomes
clearly visible when considering the relative error $|\langle A\rangle_{\mathrm{ideal}}-\langle A\rangle_{\mathrm{mit}}|/|\langle A\rangle_{\mathrm{ideal}}|$.

\section*{Conclusions}

We introduced the Heisenberg-echo symmetrization (HES) method as a
circuit-level correction to the standard gate insertion protocol.
The central result is that combining the forward evolution with its
Heisenberg echo cancels the leading error caused by the non-commutativity
of the gate noise with the ideal unitary. As a result, the leading
GI error is reduced from first order to second order in the non-commuting
noise contribution.

The mechanism can also be understood from a symmetry perspective.
Expectation values are invariant under a perfect Heisenberg echo.
In the implemented echo, however, CZ and CX gates are used as their
own inverses, as in the GI protocol, and this breaks the symmetry
in the presence of noise. Consequently, the implemented Heisenberg
echo yields an expectation value that differs from the forward evolution.
To leading order, the deviation from the ideal value has the opposite
sign to that of the forward evolution. Averaging the forward and echo
results, i.e., symmetrization, therefore cancels the leading-order
term responsible for the GI flaw.

The main practical advantage of HES is that it retains the simplicity
and broad applicability of GI without requiring access to pulse-inverse
operations. It can therefore be used when KIK or Layered-KIK cannot
be implemented directly, while remaining compatible with gate-level
interfaces. 

HES does not remove all higher-order errors. In multilayer circuits,
residual local quadratic terms and cross-layer contributions remain,
and unlike Layered-KIK, these errors cannot be systematically reduced
by making the amplification layers arbitrarily thin. HES therefore
provides a practical compromise: it substantially corrects the fundamental
flaw of GI while preserving the accessibility that makes gate insertion
attractive in the first place. Another major limitation of HES is
that it incurs no additional sampling overhead only when observables
are rank-one. Although this restricts the applicability of our approach,
we argue that whenever HES can be implemented without substantial
additional cost, it should be preferred over GI.

By construction, HES is resilient to noise drifts when executed in
small batches of shots. In each batch, all noise-amplification circuits
are executed using a small number of shots, potentially as few as
one. If the noise-drift timescale is longer than the batch timescale,
the resulting estimates are resilient to the drift. Crucially, in
HES each batch must include both the forward-evolution and Heisenberg-echo
circuits, so that the symmetrization properly cancels the leading-order
GI error term.

Since HES correctly amplifies the noise within its regime of validity,
it opens the possibility of using the recently introduced virtual
noise scaling post-processing method instead of Taylor coefficients,
thereby achieving a reduction of several orders of magnitude  in sampling
overhead in the strong-noise regime.

\bibliographystyle{unsrt}
\bibliography{Refs_CGI}

\end{document}